\documentclass[journal]{IEEEtran}
\ifCLASSINFOpdf
\else
\usepackage[dvips]{graphicx}
\fi
\usepackage{url}

\usepackage{verbatim}
\usepackage{url}
\usepackage[small]{caption}
\usepackage{graphicx}
\usepackage{amsmath}
\usepackage{bm}
\usepackage{diagbox}
\usepackage{bbding}
\usepackage{amsthm}
\usepackage{amssymb}
\usepackage{booktabs}
\usepackage[ruled]{algorithm2e}
\usepackage{mathtools}
\usepackage{multirow}
\usepackage{threeparttable}
\usepackage{color}
\usepackage{balance}
\usepackage{cite}
\usepackage{epstopdf}
\usepackage{cases}

\usepackage{xcolor}

\newcommand{\subparagraph}{}
\usepackage[explicit]{titlesec}

\begin{document}
\title{Deep Plug-and-Play Prior for Multitask Channel Reconstruction in Massive MIMO Systems}
\author{Weixiao Wan, Wei Chen, \IEEEmembership{Senior Member, IEEE}, Shiyue Wang, Geoffrey~Ye~Li,~\IEEEmembership{Fellow,~IEEE},~Bo~Ai,~\IEEEmembership{Fellow,~IEEE}
\thanks{Weixiao Wan, Wei Chen, Shiyue Wang and Bo Ai are with the State Key Laboratory of Advanced Rail Autonomous Operation, Beijing Jiaotong University, Beijing, China (email:\{weixiaowan,weich,shiyuewang,boai\}@bjtu.edu.cn).} \thanks{Geoffrey Ye Li is with the Department of Electrical and Electronic Engineering, Imperial College London, London, U.K (Email: geoffrey.li@imperial.ac.uk).}
}
\maketitle
\begin{abstract}
Scalability is a major concern in implementing deep learning (DL) based methods in wireless communication systems. Given various channel reconstruction tasks, applying one DL model for one specific task is costly in both model training and model storage. In this paper, we propose a novel unsupervised deep plug-and-play prior method for three channel reconstruction tasks in the downlink of massive multiple-input multiple-output (MIMO) systems, including channel estimation, antenna extrapolation and channel state information (CSI) feedback. The proposed method corresponding to these three channel reconstruction tasks employs a common DL model, which greatly reduces the overhead of model training and storage. Unlike general multi-task learning, the DL model of the proposed method does not require further fine-tuning for specific channel reconstruction tasks. Extensive experiments are conducted on the DeepMIMO dataset to demonstrate the convergence, performance, and storage overhead of the proposed method for the three channel reconstruction tasks.

\end{abstract}
	
\IEEEpeerreviewmaketitle
\section{Introduction}
\IEEEPARstart{M}{assive} multiple-input multiple-output (MIMO) technology improves system performance by significantly increasing the number of antennas to enhance beamforming gain and reduce inter-user interference, making it a key technology for 5G communication systems \cite{larsson2014massive}. However, with the increase in the number of antennas, the dimension of the wireless channel grows rapidly, which brings huge challenges in various channel reconstruction tasks, e.g., channel estimation \cite{8,9} and channel state information (CSI) feedback \cite{10,11}.

\par
While wireless communication systems continue to develop at a rapid pace, artificial intelligence (AI) technology has also set off a new round of technological revolution. In recent years, deep learning (DL) technology, as a branch of AI, has demonstrated breakthrough performance over classical algorithms in many fields, such as computer vision and natural language processing. AI has been applied for various tasks on the wireless physical layer in massive MIMO systems and displays excellent performance \cite{12,9905727}. AI based methods have also attracted wide attention in industry, and been considered for standardization. For example, a new study item on AI for new radio air interface has been approved in the 3rd generation partnership project (3GPP) for beyond 5G (B5G) networks \cite{6}.
	
\par
To deal with different channel reconstruction tasks for massive MIMO, various DL models have been developed. For example, to handle the channel estimation task under massive MIMO, the learned denoising-based approximate message passing (LDAMP) neural network in \cite{35} learns channel features from a large amount of training data and estimates channels, significantly outperforming compressed sensing (CS)-based methods. In \cite{34}, the channel matrix has been modeled as a two-dimensional picture and further processed using a convolutional neural network (CNN) and a denoising network to achieve high accuracy estimation of the channel. Similarly, in the antenna extrapolation task, the DNN-based antenna extrapolation for massive MIMO systems in \cite{49} extrapolates downlink CSI from a subset of downlink CSI. Antenna extrapolation via CNN has been implemented in \cite{50} in a reconfigurable intelligent surface (RIS)-assisted communication system. In \cite{51}, the neural network structure has been modified by an ordinary differential equation that describes the potential relationships between different data layers and improves the performance of antenna extrapolation. DL techniques have been first applied to CSI feedback in \cite{56}, where the DL-based CSI feedback method, namely CsiNet, exhibites excellent performance. Afterwards, a large amount of work \cite{57,59,61,63,64,65} has been explored on the basis of CsiNet, mainly in terms of multi-domain correlation extraction utilization, novel neural network structure design, and quantization method improvement to improve the performance of DL-based CSI feedback methods.


\par
DL-based methods have shown superior performance in dealing with various channel reconstruction tasks in wireless communication systems, while they bring new challenges. One prominent challenge is the increasing demand for training and storing of different neural network models designed to distinct tasks. Therefore, the existing one-to-one (AI model to communications task) solution is unscalable, and is not suitable for low cost sensors and mobile devices especially.

\par
The adaptability of DL models to different channel reconstruction tasks and scenarios calls for more investigation. Ideally, it is desired to employ a single DL model to handle multiple channel reconstruction tasks or scenarios. Therefore, some works have tried to apply multi-task learning to communication systems. The multi-task training method in \cite{9930135} improves the adaptability of CSI feedback networks to different channel scenarios; the multi-task learning-based precoding network in \cite{9597638} enhances the adaptability of the network to different signal-to-noise ratios (SNR) of the subchannels, which solves the problem of bit-error rate (BER) loss caused by imperfect CSI. However, multi-task learning usually requires labeled dataset for different tasks, which is not always available or expensive to obtain.

\par
Fortunately, there is a brilliant idea to solve the above problem, which has been applied in the field of image reconstruction. In \cite{98}, a plug-and-play (PnP) prior framework with a trainable nonlinear reaction-diffusion (TNRD) denoiser was proposed to solve image deblurring and image super-resolution problems. In \cite{73}, a highly flexible and effective CNN denoiser was trained, and then the CNN denoiser was inserted into the half-quadratic splitting (HQS) algorithm as a module to solve image restoration problems under multiple blurring kernels, including image denoising, image super-resolution and image deblurring problems. These studies have successfully applied a model to manage various image reconstruction tasks with satisfactory reconstruction results.

\par
In this paper, we propose a multitask method based on deep PnP prior framework for channel reconstruction in massive MIMO systems, which does not require labeled dataset for different channel reconstruction tasks. Specifically, we consider three channel reconstruction tasks including channel estimation, antenna extrapolation, and CSI feedback. Although these tasks are usually formulated as different optimization problems and treated by quite different methods in the literature \cite{34,35,49,50,56,57,59,61,63,64,65}, we find that they all exploit the characteristics of the wireless channel. Therefore, we formulate these three tasks as different optimization problems with a common regularization term that acts as the prior of the wireless channel. The main idea is to unroll different optimization problems into different task-specific subproblems and a common channel-specific subproblem by the variable splitting technique, and solve the common channel-specific subproblem by a common DL-based denoiser. By alternating iterative optimization of the task-specific subproblem and the channel-specific subproblem, we could handle the three different tasks with only a single DL model. The proposed method greatly reduces the overhead of model training and storage. Furthermore, the plug-and-play methods are essentially unsupervised learning models. This means that the same model is not only reused, but also trained unsupervised. Different from general multi-task learning \cite{9597638,9930135}, the DL model of the proposed method does not require labeled dataset and further fine-tuning for specific channel reconstruction tasks.

\par
The remainder of this paper is organized as follows. Section \ref{sec:Model} introduces the system model. Section \ref{method} presents the proposed method. Section \ref{Experimentel Results} provides the experimental results and Section \ref{sec:Conclusions} draws the conclusion.
	
\par
$\mathit{Notation:}$ Bold uppercase $\mathbf{A}$ and bold lowercase $\mathbf{a}$ denote a matrix and a column vector, respectively, non-bold letter $a$ and $A$ are scalars, blackboard bold letter $\mathbb{A}$ is a set and $\mathbb{E}$ refers specifically to the expectation; Caligraphic letter $\mathcal{A}$ is the mapping, $ \left\| \mathbf{a}\right\|_{2} $ is the 2-norm of a vector, $ \left\| \mathbf{A} \right\|_{2} $ is the Frobenius norm of a matrix; $\mathbf{A}^{\text{T}}$ , $\mathbf{A}^{-1}$, $\mathbf{A}^{\text{H}}$ are the transpose, inverse, and hermitian of $\mathbf{A}$; $ \circ $ represents the hadamard product operator, i.e., the element-by-element multiplication.

\begin{figure*}[!tb]l
\centering
\includegraphics[scale=0.5]{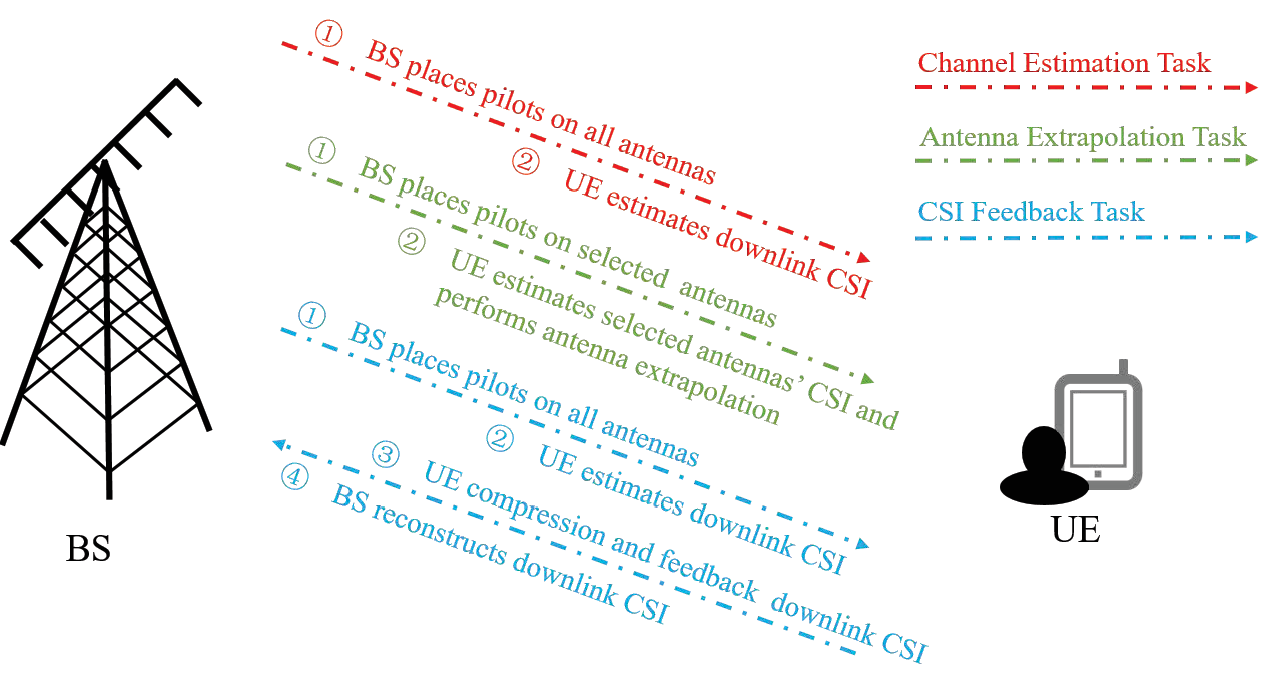}
\caption{Schematic diagram of three channel reconstruction tasks in massive MIMO systems.}
\label{fig:04:01}
\end{figure*}

\section{Background and System Model} \label{sec:Model}
We consider a single-cell massive MIMO orthogonal frequency division multiplexing (OFDM) system operating in the frequency division duplex (FDD) mode. The BS is equipped with $N_t$ antennas with uniform linear array (ULA) arrangement, and the user equipment (UE) is equipped with a single antenna due to limited hardware cost and power consumption. Assume that the channel between the UE and the BS consists of $L$ paths. According to the spatially correlated physical channel model in \cite{85}, the downlink channel vector ${\mathbf{h}_n}\in {\mathbb{C}^{N_t}}$ between the BS and the UE on the $n$th subcarrier can be expressed as
\begin{equation}\label{eq:02:01}	
{{\mathbf{h}}_{n}}=\underset{l=1}{\mathop{\overset{L}{\mathop{\sum }}\,}}\,{{\alpha }_{l}}{{e}^{-j2\pi {{f}_{n}}{{\tau }_{l}}+j{{\phi }_{l}}}}\mathbf{a}\left( {{\theta }_{l,n}} \right),
\end{equation}
where ${f}_{n}$ denotes the subcarrier frequency corresponding to the $n$th subcarrier, ${\alpha }_{l}$, ${\phi }_{l}$, and ${\tau }_{l}$ denote the attenuation coefficient, phase shift, and time delay of the $l$th path, respectively, and ${\theta }_{l,n}$ denotes the angle of arrival (AoA) for the $n$th subcarrier and the $l$th path. In addition, $\mathbf{a}\left({{\theta }_{l,n}}\right)$, the steering vector of the antenna array, is defined as
\begin{equation}\label{eq:02:02}
\mathbf{a}\left( {{\theta }_{l,n}} \right)\!=\!{{\left[ 1, {{e}^{-j\frac{2\pi d{{f}_{n}}}{c}\sin {{\theta }_{l,n}}}}, \cdots , {{e}^{-j\frac{2\pi d{{f}_{n}}}{c}\left( {N_t}-1 \right)\sin {{\theta }_{l,n}}}} \right]}^\text{T}},
\end{equation}
where $d$ denotes the antenna spacing and $c$ is the speed of light. In this paper, we consider three downlink channel reconstruction tasks, i.e., channel estimation, antenna extrapolation, and CSI feedback. An illustration of the three tasks is shown in Fig.~\ref{fig:04:01}.
	
\subsection{Channel Estimation}
\par
As shown in Fig.~\ref{fig:04:01}, downlink channel estimation consists of two steps. Firstly, the BS sends allocated pilots for all antennas in different OFDM symbols. Then the UE estimates the downlink channel based on the known pilots and the received signals. Assuming that the OFDM system has $N_s$ subcarriers, the complete downlink CSI $\mathbf{H} \in {\mathbb{C}}^{{N_s}\times{N_t}}$ in the spatial-frequency domain can be expressed as
\begin{equation}\label{eq:02:03}	
\mathbf{H}={\left[ {\mathbf{h}}_{1},\ {\mathbf{h}}_{2},\ \cdots ,\ {\mathbf{h}}_{N_s} \right]}^\text{T}.
\end{equation}
Then the relationship between the transmitted signal and the received signal can be expressed as
\begin{equation}\label{eq:02:04}	
\mathbf{Y}=\mathbf{H}\circ \mathbf{X}+\mathbf{W},
\end{equation}
where $\mathbf{X} \in {{\mathbb{C}}^{{N_s}\times{N_t}}}$ and $\mathbf{Y} \in {{\mathbb{C}}^{{N_s}\times{N_t}}}$ denote the transmitted signal and the received signal, respectively, and $\mathbf{W}\in {{\mathbb{C}}^{{N_s}\times{N_t}}}$ denotes additive white Gaussian noise with mean equal to 0 and variance equal to ${{\sigma }^{2}}$.
\par
To conduct channel estimation, the pilots are usually placed sparsely in a grid of the spatial-frequency domain, and the length of pilots is usually proportional to the number of antennas. For a specific pilot pattern, i.e. positions for pilot signal, the received signal ${{\mathbf{Y}}_{p}}\in {{\mathbb{C}}^{{N_{sp}}\times {N_{t}}}}$ at the UE can be expressed as
\begin{equation}\label{eq:02:05}	
{{\mathbf{Y}}_{p}}={{\mathbf{H}}_{p}}\circ {{\mathbf{X}}_{p}}+{{\mathbf{W}}_{p}},
\end{equation}
where $\mathbf{X}_{p} \in {\mathbb{C}}^{{N_{sp}} \times {N_{t}}}$ denotes the transmitted pilots, and $\mathbf{H}_{p}\in {\mathbb{C}}^{N_{sp}\times {N_{t}}}$ is the channel response at the pilot location. Here, $N_{sp}$ is the number of pilot symbols placed along the frequency domain. The relationship between the CSI at pilot positions and the complete downlink CSI $\mathbf{H}$ is given by
\begin{equation}\label{eq:02:06}{{\mathbf{H}}_{p}}=\mathcal{P}\left( \mathbf{H} \right),
\end{equation}
where $\mathcal{P}:{\mathbb{C}}^{{N_s}\times {N_t}} \to {\mathbb{C}}^{N_{sp}\times N_{t}}$ denotes the downsampling operator corresponding to the pilot pattern. In fact, in massive MIMO systems, the pilot positions of different users are usually orthogonal in order to prevent pilot contamination as much as possible. In addition, according to the variation of channel quality, the density of pilots could be adjusted to obtain satisfactory channel estimation performance. Therefore, the pilot patterns used by different users in different communication scenarios could be different in reality, where multiple pilot patterns are configured to fit different communication scenarios.
	
\par
To sum up, the purpose of downlink channel estimation is to estimate the complete downlink CSI $\mathbf{H}$ with the known pilots $\mathbf{X}_{p}$ and the received signal $\mathbf{Y}_{p}$. The optimization problem can be expressed as
\begin{equation}\label{eq:02:07}
\min _{\mathbf{H}}\left\|\mathbf{Y}_{p}-\mathcal{P}(\mathbf{H}) \circ \mathbf{X}_{p}\right\|_{F}^{2}+\lambda \mathcal{J}(\mathbf{H}),
\end{equation}
where $\mathcal{J}\left(\cdot\right)$ denotes a regularization term that implicitly captures the downlink CSI characteristics and $\lambda$ is a positive regularization parameter that controls the impact of the regularization term.

\par
Least-squares (LS) estimation and linear minimum mean-squared error (LMMSE) estimation are two classical channel estimation methods \cite{103}. LS estimation is known for its simplicity and convenience, but its estimation accuracy is suboptimal. LMMSE estimation usually achieves enhanced accuracy, but requires noise variance and second-order channel statistics such as the cross-correlation matrix between the real channel coefficients and the channel coefficients at pilots positions as a prior information. Without those information, the LMMSE method, which is computationally complex, cannot be applied in practice, but can be used as a performance upper bound.

The formulas in \cite{103} are given as follows. The LS estimated channel at pilot positions is obtained by solving the optimization problem below
\begin{equation}
    \mathbf{\hat{h}}_p^{\text{LS}} = \mathop{\text{argmin}}\limits _{\mathbf{\hat{h}}_p} \| \mathbf{y}_p -\mathbf{\hat{h}}_p \circ \mathbf{x}_p \|_2^2,
\end{equation}
where $\mathbf{x}_p, \mathbf{y}_p$ and $\mathbf{\hat{h}}_p$ are vectorized $\mathbf{X}_p, \mathbf{Y}_p$ and $ \mathbf{\hat{H}}_p$, respectively. $\mathbf{\hat{H}}_p$ is the estimated $\mathbf{H}_p$.
The LMMSE estimated channel is obtained by multiplying the LS estimation at pilots positions with a filtering matrix
\begin{equation}
    \mathbf{\hat{h}}^{\text{LMMSE}} = \mathbf{A}_{\text{LMMSE}} \mathbf{\hat{h}}_p^{\text{LS}},
\end{equation}
which is obtained by solving a minimization problem
\begin{equation}
\begin{array}{rl}
     \mathbf{A}_{\text{LMMSE}}
     & = \mathop{\text{argmin}}\limits_{\mathbf{A}} \| \mathbf{h} - \mathbf{A} \mathbf{\hat{h}}_p^{\text{LS}} \|_2^2 \\
     & = \mathbf{R}_{\mathbf{h},\mathbf{h}_p} ( \mathbf{R}_{\mathbf{h}_p,\mathbf{h}_p} + \sigma^2 \mathbf{I} ) ^{-1},
\end{array}
\end{equation}
where $\mathbf{h}$ and $ \mathbf{h}_p$ are vectorized $\mathbf{H}$ and $\mathbf{H}_p$. $\mathbf{R}_{\mathbf{h}, \mathbf{h}_p} = \mathbb{E}(\mathbf{h} \mathbf{h}_p^H)$ is the cross-correlation matrix between $\mathbf{h}$ and $\mathbf{h}_p$. $\mathbf{R}_{\mathbf{h}_p, \mathbf{h}_p} = \mathbb{E}(\mathbf{h}_p \mathbf{h}_p^H)$ is the channel autocorrelation matrix at pilot positions.

DL-based channel estimation methods can improve performance \cite{34,35,36,103}. However, the existing methods usually consider a specific pilot pattern, while the corresponding DL model also needs to be replaced by another model if the pilot pattern changes. In the 5G systems, different pilot patterns for various communication scenarios can be configured \cite{3GPP1,3GPP2,3GPP3,3GPP4}, which brings huge model training and storage overhead for AI models.
		
\subsection{Antenna Extrapolation}
\par
In massive MIMO systems, the increase in the number of antennas brings higher spatial freedom and multiplexing capability to the system, which results in a huge system performance gain. However, the increase in the number of antennas also brings a large pilot overhead, which is proportional to the number of antennas. To reduce the pilot overhead required for channel estimation, some work has attempted to exploit the correlation between antennas to obtain the CSI of all antennas via only the CSI of selected antennas, known as antenna extrapolation \cite{49,50,51}. However, unlike in the frequency domain, the relationships between the channels in the antenna domain do not have explicit mathematical expressions, even though the channels are highly correlated \cite{49}. Antenna extrapolation can significantly reduce pilot overhead, as pilots can be placed only on selected antennas \cite{39}.

\par
As shown in Fig.~\ref{fig:04:01}, in the downlink antenna extrapolation task, the BS places pilots on the selected antennas, and the UE estimates the CSI of the selected antennas. The antenna extrapolation is conducted, which can be considered as a post-processing of downlink channel estimation. Assume that the BS selects $\bar{N}_t$ antennas for antenna extrapolation according to an arbitrary antenna selection pattern, then the sampling rate can be defined as $r = \bar{N}_t/N_t$. The relationship between the downlink CSI, $\tilde{\mathbf{H}}\in{{\mathbb{C}}^{{N_s}\times{\bar{N}_t}}}$, of the selected antennas and the complete CSI, $\mathbf{H}\in{{\mathbb{C}}^{{N_s}\times{N_t}}}$, of all antennas can be expressed as
\begin{equation}\label{eq:02:11}	
\tilde{\mathbf{H}}=\mathcal{A}\left( \mathbf{H} \right),
\end{equation}
where $\mathcal{A}:{\mathbb{C}}^{{N_s}\times {N_t}} \to {\mathbb{C}}^{N_{s}\times \bar{N}_{t}}$ denotes the antenna selection pattern. Here, with reference to the assumptions in \cite{49}, the downlink CSI of selected antennas has been perfectly estimated. We consider the selection of antennas with uniform intervals, although other selection patterns are admitted. Note that different antenna selection patterns will result in different antenna extrapolation performance. The antenna extrapolation can also be formulated as the following optimization problem
\begin{equation}\label{eq:02:12}	
\min_{\mathbf{H}}\left\|\tilde{\mathbf{H}}-\mathcal{A}\left( \mathbf{H} \right)\right\|_{F}^{2}+\lambda \mathcal{J}(\mathbf{H}),
\end{equation}
where $\mathcal{J}\left(\cdot\right) $ denotes the regularization term that implicitly captures the downlink CSI characteristics.

\par
To solve the above optimization problem, in \cite{49},  the proposed DL model is trained for a specific antenna selection pattern. If the antenna selection pattern is changed, the DL model would no longer be valid. This drawback has also been noticed and discussed in the literature \cite{48,50,51}.

\subsection{CSI Feedback} \label{CSI feedback}
\par
In FDD massive MIMO systems, the BS needs accurate downlink CSI \cite{52} to perform precoding and beamforming. As shown in Fig.~\ref{fig:04:01}, once the UE obtains the downlink CSI $\mathbf{H}$, some form of information needs to be fed back to the BS. To reduce feedback overhead, a two-dimensional discrete Fourier transform (DFT) is employed to transform the CSI into the angular-delay domain, which leads to
\begin{equation}\label{eq:02:14}	
\breve{\mathbf{H}}=\mathbf{F}_{s}\mathbf{H}\mathbf{F}_{t},
\end{equation}
where $\mathbf{F}_{s}$ and $\mathbf{F}_{t}$ are the DFT matrices of dimension $N_s \times N_s$ and $N_t \times N_t$, respectively. As the large CSI coefficients in the angular-delay domain occupy only a small portion and the other coefficients are close to zero, the angular-delay domain downlink CSI $\breve{\mathbf{H}}$ is sparse. Moreover, since the time delay between multipath arrivals at the UE is in a finite period, almost all large coefficients are in the first $\bar{N}_s$ rows of $\breve{\mathbf{H}}$. Therefore, the first $\bar{N}_s$ rows of $\breve{\mathbf{H}}$, denoted as $\bar{\mathbf{H}}\in\mathbb{C}^{\bar{N}_s \times N_t}$, are fed to the BS and the rest of the rows are simply ignored to reduce the feedback overhead in \cite{56}.
	
\par
To further reduce the feedback overhead, various DL-based methods have been proposed to compress the angular-delay domain downlink CSI. The mostly considered DL architecture for CSI compression is the autoencoder, where the UE and the BS use a DL-based encoder and a DL-based decoder, respectively \cite{56,57,59,60,61}. The encoder DL model at the UE maps $\bar{\mathbf{H}}$ to a low-dimensional compressed space, while the decoder DL model at the BS maps the received feedback information to the original dimension to construct $\bar{\mathbf{H}}$. However, the two-sided DL model at both the UE and the BS has several drawbacks. First, the use of DL technique comes at the considerable cost of computational complexity and model storage overhead, which severely hinders its use in low-cost UE, such as IoT devices. Second, joint training of the two-sided DL model requires consensus and collaborations between the BS and the UE, which is not easy to achieve in practice considering the standardization issues.

\par
To overcome these shortcomings, a one-sided CSI feedback framework is proposed in \cite{chen2022csi}, where only the BS-side requires the deployment of DL models. On the UE side, the CSI is simply compressed by linear mapping, which can significantly reduce the computational burden on the UE side and eliminate the need for collaborations.
In the one-sided CSI feedback framework, the compression process at the UE side can be expressed as
\begin{equation}\label{eq:03:01}	
\mathbf{y}=\mathbf{A}\bar{\mathbf{h}}+\mathbf{w},
\end{equation}
where $\bar{\mathbf{h}}\in \mathbb{R}^N$ consists of the real and imaginary parts of the truncated angular-delay domain CSI $\bar{\mathbf{H}}$ spliced after vectorization, $N=2\bar{N}_s N_t$, $\mathbf{y}\in \mathbb{R}^M$, $\mathbf{A}\in \mathbb{R}^{M \times N}$ denote the compressed CSI vector and the linear projection matrix, respectively, and $\mathbf{w}\in \mathbb{R}^M$ denotes the quantization noise caused by the quantization process. The dimension $M$ of the compressed CSI vector $\mathbf{y}$  is usually much smaller than the dimension $N$ of the CSI vector $\bar{\mathbf{h}}$, so the compression ratio (CR) defined as $M/N$\ is much lower than 1. The CSI reconstruction problem at the BS side can be formulated as
\begin{equation}\label{eq:03:03}		
\min_{\bar{\mathbf{h}}}\ \left\|\mathbf{y}-\mathbf{A}\bar{\mathbf{h}}\right\|_{2}^{2}+\lambda \mathcal{J}\left( \bar{\mathbf{h}}\right),
\end{equation}
where $\mathcal{J}\left(\cdot\right)$ denotes the regularization term that implicitly captures the angular-delay properties of the channel response by using the CSI vector $\bar{\mathbf{h}}$ as a priori information.
	

\section{Multi-task Deep Plug-and-Play Prior}\label{method}
In this section, we propose a multi-task deep PnP prior method that employs a common DL model to handle different tasks in massive MIMO systems including channel estimation, antenna extrapolation, and CSI feedback. We first introduce the deep PnP prior method to solve linear inverse problems and the design perspectives of DL-based denoiser. Then we provide the details about how the three communication tasks formulated in Section II can be solved.

\subsection{Deep PnP prior}\label{deep PnP prior}
The three communication tasks formulated in Section II boil down into linear inverse problems. Mathematically, the so-called linear inverse problem is to recover the unknown signal $\mathbf{x}\in\mathbb{R}^{N}$ from the observed data $\mathbf{y}\in\mathbb{R}^{M}$ by a known linear observation mapping $\mathcal{T}\left(\cdot \right) $, and the observed data are usually corrupted by some noise $\mathbf{n}\in\mathbb{R}^{M}$, given by
\begin{equation}\label{eq:02:21}\mathbf{y}=\mathcal{T}\left(\mathbf{x} \right)+\mathbf{n}.
\end{equation}
The recovery of $\mathbf{x}$ from observation data $\mathbf{y}$ can be formulated as the following optimization problem
\begin{equation}\label{eq:02:22}\min_{\mathbf{x}}\left\|\mathbf{y}-\mathcal{T}\left(\mathbf{x} \right)\right\|_{2}^{2}+\lambda \mathcal{J}(\mathbf{x}),
\end{equation}
where $\left\|\mathbf{y}-\mathcal{T}\left(\mathbf{x} \right)\right\|_{2}^{2}$ is the  data-fidelity term that captures the difference between the recovered signal and the original signal, $\mathcal{J}\left(\cdot \right) $ is the regularization term that captures prior signal information, and $\lambda$ is the positive regularization parameter that promotes the balance between the data-fidelity term and the regularization term. In specific, for the three communication tasks described in Section \ref{sec:Model}, the data-fidelity terms corresponding to channel estimation, antenna extrapolation and CSI feedback are $\left\|\mathbf{Y}_{p}-\mathcal{P}(\mathbf{H}) \circ \mathbf{X}_{p}\right\|_{F}^{2}$, $\left\|\tilde{\mathbf{H}}-\mathcal{A}\left( \mathbf{H} \right)\right\|_{F}^{2}$ and $\left\|\mathbf{y}-\mathbf{A}\bar{\mathbf{h}}\right\|_{2}^{2}$, respectively. We would like to emphasize that implicit prior information on the channel matrix contained by regularization term in three tasks are highly complex, and it is challenging to create a prior for real-world channel datasets using analytical methods.
	
\par
In order to solve the above optimization problem, variable splitting algorithms such as alternating direction method of multipliers (ADMM) and HQS could be employed. As HQS is concise and converges quickly \cite{73}, we use HQS to decouple the data-fidelity term from the regularization term. By introducing an auxiliary variable $\mathbf{z}$, the optimization problem in $\left( \ref{eq:02:22}\right)$ can be rewritten as
\begin{equation}\label{eq:04:01}
\begin{aligned}
\min_{\mathbf{x}}\quad&\left\|\mathbf{y}-\mathcal{T}\left(\mathbf{x} \right)\right\|_{2}^{2}+\lambda \mathcal{J}(\mathbf{z})\\
s.t.\quad &\ \mathbf{z}=\mathbf{x}.
\end{aligned}
\end{equation}
$\left( \ref{eq:04:01}\right)$ is then solved by minimizing the following surrogate problem
\begin{equation}\label{eq:04:02}\mathcal{L}_{\rho}\left(\mathbf{x},\ \mathbf{z}\right) =\left\|\mathbf{y}-\mathcal{T}\left(\mathbf{x} \right)\right\|_{2}^{2}+\lambda \mathcal{J}\left( \mathbf{z}\right)+\rho\left\|\mathbf{z}-\mathbf{x}\right\|_{2}^{2},
\end{equation}
where $\left\|\mathbf{z}-\mathbf{x}\right\|_{2}^{2}$ is the constraint term and $\rho$ is the positive penalty parameter. By gradually increasing $\rho$, the solution of $ \left( \ref{eq:04:02}\right) $ will get close to the solution of $ \left( \ref{eq:04:01}\right) $ . The optimization problem in $ \left( \ref{eq:04:02}\right) $  can be solved by iteratively solving the subproblems with variables $\mathbf{z}$ and $\mathbf{x}$ in turn, which is expressed as
\begin{subequations}\label{eq:04:03}
\begin{numcases}{}
\mathbf{x}^{t+1}:=\arg\min_{\mathbf{x}}\left\|\mathbf{y}-\mathcal{T}\left(\mathbf{x} \right)\right\|_{2}^{2}+\rho\left\|\mathbf{z}^{t}-\mathbf{x}\right\|_{2}^{2},\label{eq:04:03a}\\
\mathbf{z}^{t+1}:=\arg\min_{\mathbf{z}}\mathcal{J}\left(\mathbf{z}\right)+\frac{1}{2\left(\sqrt{\frac{\lambda}{2\rho}} \right) ^{2}}\left\|\mathbf{z}-\mathbf{x}^{t+1}\right\|_{2}^{2}.\label{eq:04:03b}
\end{numcases}
\end{subequations}

\par
The subproblem in $ \left( \ref{eq:04:03a}\right) $ leads to the proximal solution, $\mathbf{x}^{t+1}$, which usually has a closed-form solution owing to linear mapping $\mathcal{T}\left(\cdot \right) $. However, computing subproblem $\left(\ref{eq:04:03b}\right)$ is challenging as it requires explicit knowledge of the prior $\mathcal{J}\left(\cdot\right)$, even when we has a great approximation of $\mathcal{J}\left(\cdot\right)$, it still be extremely difficult to carry out analytically. Fortunately, it is invariant of the task and the data \cite{10056957}. In particular, it can be seen as the Maximum A Posteriori (MAP) estimation problem, where the prior of $\mathbf{z}$ is $e^{-\mathcal{J}\left(\mathbf{z}\right)}$ and $\mathbf{z}$ is corrupted by a Gaussian noise with variance $ \sigma^{2}=\frac{\lambda}{2\rho}$\cite{10056957}. Thus, a denoiser can be used to solve $ \left( \ref{eq:04:03b}\right) $ implicitly, which can be rewritten as follows
\begin{equation}\label{eq:04:04}
\mathbf{z}^{t+1}=Denoiser\left(\mathbf{x}^{t+1},\ \sigma^{2}=\frac{\lambda}{2\rho} \right).
\end{equation}
Noted that $\left(\ref{eq:04:03b}\right)$ is essentially the proximal mapping with respect to (a scaled) $\mathcal{J}\left(\cdot\right)$.
\par
Significantly, the parameters, $\rho$ and $\lambda $, are involved in the whole alternating iterative optimization process, and the setting of the two parameters affects convergence. To ensure that $\mathbf{x}^{t+1}$ and $\mathbf{z}^{t+1} $ converge to a fixed point, $\rho$, i.e., the weight of the constraint term, needs to keep getting larger during the iterative process. The increase of $\rho$ can be also viewed as the decrease of the noise variance in (\ref{eq:04:04}), and $\mathbf{x}^ {t+1}$ and $ \mathbf{z}^{t+1} $ will gradually converge to the true value. The process on solving linear inverse problem via PnP prior is summarized in the algorithm \ref{alg:04:01}.

\begin{algorithm}[t]
\SetAlgoLined
\LinesNumbered
\DontPrintSemicolon
\caption{Solving Linear Inverse Problem via PnP prior.}
\label{alg:04:01}
\KwIn{Observed data $\mathbf{y}$, linear observation mapping $\mathcal{T}\left(\cdot \right)$, regularization parameters $\lambda$, penalty factor $\rho$, scaling factor $\alpha$, number of iterations $N$}
\KwOut{Reconstructed signal $\hat{\mathbf{x}}$}
Initialize input data $\mathbf{z}^{1}$\;
\For{$t = 1:N$}{
Compute $\mathbf{x}^{t+1}$ by $ \left( \ref{eq:04:03a}\right) $  \;
Compute $\mathbf{z}^{t+1}$ by $ \left( \ref{eq:04:04}\right)  $\;
Update penalty factor $\rho^{t+1}=\alpha\rho^{t}$;	}
return $\mathbf{x}^{N+1}$
\end{algorithm}

\begin{figure}[!tb] 
\centering
\includegraphics[scale=0.6]{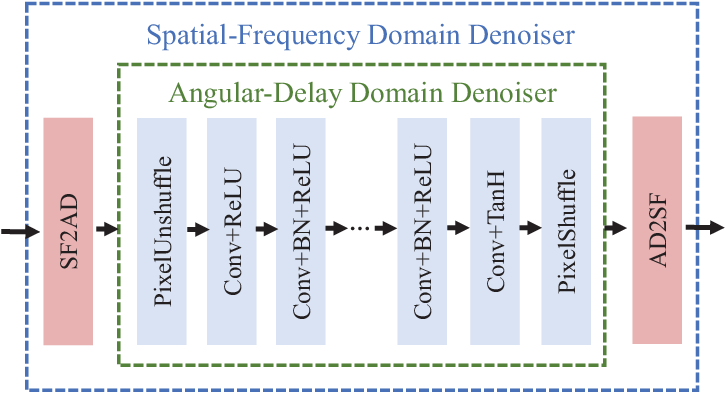}
\caption{Schematic diagram of the denoising network structure.}
\label{fig:04:02}
\end{figure}

\begin{figure}[!tb] 
\centering
\includegraphics[scale=0.6]{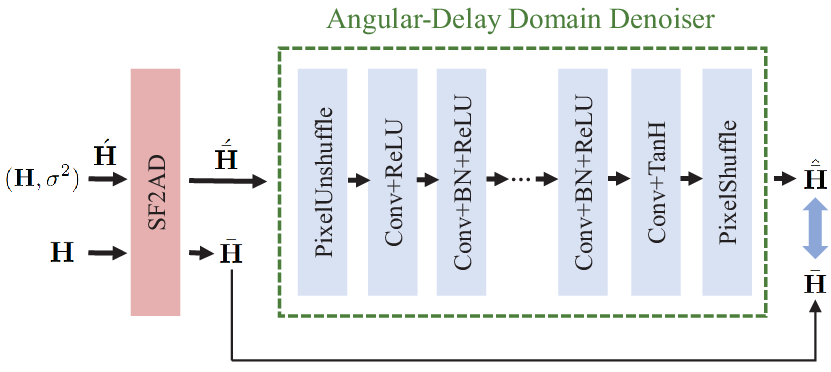}
\caption{Training diagram of the denoising network.}
\label{fig:04:13}
\end{figure}

\par
According to the optimization process discussed above, the noise variance in $(\ref{eq:04:04}) $ changes continuously during the iterative process, which means that the denoiser needs to have the ability to handle a varying noise level. In this paper, we design a DL-based denoiser that can be used simultaneously for the three channel reconstruction tasks, i.e., channel estimation, antenna extrapolation, and CSI feedback. In specific, the channel estimation problem in (\ref{eq:02:07}) and the antenna extrapolation problem in (\ref{eq:02:12}) directly deal with the downlink CSI $\mathbf{H}$ in the spatial-frequency domain, while the CSI feedback problem in (\ref{eq:03:03}) deals with the tailored CSI $\breve{\mathbf{H}}$ in the angular-delay domain, which exploits the limited time delay of multi-path to reduce redundancy. To deal with the channel in different domains, we consider a denoising network with a structure shown in Fig~\ref{fig:04:02} and its training process shown in Fig~\ref{fig:04:13}.
The DL-based denoiser is trained in the angular-delay domain but works in both the angular-delay and the spatial-frequency domains.
\par
In the training stage, the clean spatial-frequency domain CSI $\mathbf{H}$ is first added to the noise $\sigma^2$ to generate the noisy spatial-frequency domain CSI $\acute{\mathbf{H}}$ whose SNR is uniformly distributed in the range of $0\sim40$dB. Then the truncated noisy angular-delay domain CSI $\acute{\bar{\mathbf{H}}}$ and the truncated clean angular-delay domain CSI $\bar{\mathbf{H}}$ after SF2AD module become the input of the denoising network and its training labels, respectively.
The DL model structure is shown in the blue box in Fig.~\ref{fig:04:02}. The first layer of the denoiser is the pixel unshuffle layer, followed by several convolutional layers to extract features, and finally restored to the original dimension by the pixel shuffle layer. Considering the balance between the computational complexity of the network and the denoising performance, we set up 8 convolutional layers in the middle convolutional operation, and each convolutional layer has 48 convolutional kernels. The denoiser can be expressed as
\begin{equation}\label{eq:04:05}
\hat{\bar{\mathbf{H}}}=Denoiser\left(\acute{\bar{\mathbf{H}}},\ \sigma^{2};\ \Theta \right),
\end{equation}
where $\Theta $ denotes the set of parameters of the DL-based denoiser and $\sigma^{2}$ denotes the noise variance. To learn the model parameters $\Theta$ from a noisy CSI dataset, the loss function for training this denoiser is given by
\begin{equation}\label{eq:03:17}
\mathcal{L}\left(\Theta \right)=\frac{1}{T} \sum_{j=1}^{T}\frac{\left\| \bar{\mathbf{H}}_{j}-Denoiser\left(\acute{\bar{\mathbf{H}}}_{j},\ \sigma^{2};\ 	\Theta \right)\right\| _{F}^{2}}{\left\| \bar{\mathbf{H}}_{j}\right\|_{F}^{2}},
\end{equation}
where $T$ is the total number of samples in the training set, subscript $j$ denotes the $j$th sample in the training set, and $\left\|\cdot\right\| _{F}$ denotes the Frobenius norm. Note that the common DL-based denoisor can be trained off-line without supervision and reused for three different tasks in the inference stage.

The SF2AD module is applied before the denoisor, which converts the spatial-frequency CSI into the angular-delay domain CSI $\mathbf{H}$ by 2D DFT, and truncates the part with small elements to obtain the truncated angular-delay domain CSI $\bar{\mathbf{H}}$ for denoising, which can be expressed as
\begin{equation}\label{SF2AD}
\bar{\mathbf{H}}=SF2AD\left(\mathbf{H}\right)=f_{cropping}\left(DFT\left(\mathbf{H}\right)\right),
\end{equation}
where $f_{cropping}\left(\cdot\right)$, a cropping function, crops off the part of the angular-delay domain CSI $\breve{\mathbf{H}}$ with small elements. The AD2SF module feeds zeros back into the truncated angular-delay domain CSI after denoising, and then converts it to the spatial-frequency domain by 2D IDFT. The process is expressed as
\begin{equation}\label{AD2SF}
\mathbf{H}=AD2SF\left(\bar{\mathbf{H}}\right)=IDFT\left(f_{padding}\left(\bar{\mathbf{H}}\right)\right),
\end{equation}
where $f_{padding}\left(\cdot\right)$ denotes the zero-padding function that zero-pads the cropped part to the original dimension. By using the SF2AD and the AD2SF modules in the inference stage, the DL-based denoiser originally designed for the CSI in angular-delay domain can be reused to deal with CSI in spatial-frequency domain.

\par

\subsection{Channel Estimation via Deep PnP prior }

This subsection describes the proposed algorithm, i.e., PPPCE, for the downlink channel estimation task. According to (\ref{eq:02:07}) and (\ref{eq:04:02}), the subproblem in (\ref{eq:04:03a}) can be rewritten as
\begin{equation}\label{eq:04:06}
\displaystyle
\mathbf{H}^{t+1}:=\arg\min_{\mathbf{H}}\left\|\mathbf{Y}_{p}-\mathcal{P}(\mathbf{H}) \circ \mathbf{X}_{p}\right\|_{F}^{2}+\rho\left\|\mathbf{Z}^{t}-\mathbf{H}\right\|_{F}^{2},
\end{equation}
which has a closed-form solution as following
\begin{equation}\label{eq:04:09}
h_{ij}^{t+1}=\begin{cases}
\frac{x_{ij}y_{ij}+\rho z_{ij}^{t}}{x_{ij}^{2}+\rho},&\left(i,\ j\right)\in \mathbb{P}\\
z_{ij}^{t},&\left(i,\ j\right)\in \bar{\mathbb{P}},
\end{cases}
\end{equation}
where $h_{ij}$, $x_{ij}$, $y_{ij}$, and $z_{ij}^{t}$ denote the $(i,\ j)$-th element in the CSI $\mathbf{H}$, corresponding element in the transmitted pilots $\mathbf{X}_{p}$, corresponding element in the received pilots $\mathbf{Y}_{p}$, and the $(i,\ j)$-th element in the denoising result in the $t$th iteration $\mathbf{Z}^{t} $, respectively. The sets $\mathbb{P}$ and $\bar{\mathbb{P}}$ denote the pilot and non-pilot positions, respectively. The proof of (\ref{eq:04:09}) can be found in Appendix \ref{proof1}.
	
\par
Subproblem (\ref{eq:04:03b}) can be solved by the DL-based denoiser described in the previous subsection. The proposed plug-and-play channel estimation algorithm, i.e., PPPCE, is shown in Algorithm \ref{alg:04:02}, where the LS estimation is used for initialization. Note that PPPCE can be used for arbitrary pilot patterns since the DL-based denoiser works on the angular-delay domain and is decoupled from the configuration of the pilot patterns.

\begin{algorithm}[tb!]
\SetAlgoLined
\LinesNumbered
\DontPrintSemicolon
\caption{PPPCE.}
\label{alg:04:02}
\KwIn{Transmitted pilots $ \mathbf{X}_{p} $, received pilots $ \mathbf{Y}_{p} $, pilot pattern $ \mathcal{P}\left( \cdot \right) $, regularization parameters $\lambda$, penalty factor $\rho$, scaling factor $\alpha$, number of iterations $N$}
\KwOut{Estimated downlink CSI $ \hat{\mathbf{H}}$}
Initialization by LS estimation, get $\mathbf{Z}^{1}$\;
\For{$t = 1:N$}{
Compute $\mathbf{H}^{t+1}$ by (\ref{eq:04:09})\;
Convert $\mathbf{H}^{t+1}$ to the angular-delay domain and truncate it by SF2AD module\;
Compute $\mathbf{Z}^{t+1}$ by (\ref{eq:04:05})\;
Convert $\mathbf{Z}^{t+1}$ to the spatial-frequency domain by AD2SF module with zero padding\;
Update penalty factor $\rho^{t+1}=\alpha\rho^{t}$;	}
return $\mathbf{Z}^{N+1}$
\end{algorithm}

\subsection{Antenna Extrapolation via Deep PnP prior}
\par
This subsection describes the proposed algorithm, i.e., PPPAE, for the downlink antenna extrapolation task. According to (\ref{eq:02:12}), (\ref{eq:04:01}), and (\ref{eq:04:02}), subproblem (\ref{eq:04:03a}) for the antenna extrapolation task can be expressed as
\begin{equation}\label{eq:04:10}
\mathbf{H}^{t+1}:=\arg\min_{\mathbf{H}}\left\|\tilde{\mathbf{H}}-\mathcal{A}\left( \mathbf{H} \right)\right\|_{F}^{2}+\rho\left\|\mathbf{Z}^{t}-\mathbf{H}\right\|_{F}^{2}.
\end{equation}
Its closed-form solution is given by	
\begin{equation}\label{eq:04:13}
h_{ij}^{t+1}=\begin{cases}
\frac{\tilde{h}_{ij}+\rho z_{ij}^{t}}{1+\rho},&\left(i,\ j\right)\in \mathbb{A}\\
z_{ij}^{t},&\left(i,\ j\right)\in \bar{\mathbb{A}},
\end{cases}
\end{equation}	
where $h_{ij}$, $\tilde{h}_{ij}$, and $z_{ij}^{t}$ denote the $(i,\ j)$-th elements in the full CSI $ \mathbf{H} $, corresponding element in observed CSI $ \tilde{\mathbf{H}}$ and the $(i,\ j)$-th elements in the result of the denoiser in the $t$th iteration $\mathbf{Z}^{t} $, respectively. The sets $\mathbb{A}$ and $\bar{\mathbb{A}}$ denote the selected antenna and unselected antenna positions, respectively, and the proof of (\ref{eq:04:13}) can be referred to in Appendix \ref{proof2} .
\begin{algorithm}[tb!]
\SetAlgoLined
\LinesNumbered
\DontPrintSemicolon
\caption{PPPAE.}
\label{alg:04:03}
\KwIn{The selected antennas' downlink CSI $ \tilde{\mathbf{H}}$, antenna selection pattern $ \mathcal{A}\left( \cdot \right) $, regularization parameters $\lambda$, penalty factor $\rho$, scaling factor $\alpha$, number of iterations $N$}
\KwOut{Extrapolated downlink CSI $ \hat{\mathbf{H}}$}
Initialization by spline interpolation, get $\mathbf{Z}^{1}$\;
\For{$t = 1:N$}{
Compute $\mathbf{H}^{t+1}$ by (\ref{eq:04:13})\;
Convert $\mathbf{H}^{t+1}$ to the angular-delay domain and truncate it by SF2AD module\;
Compute $\mathbf{Z}^{t+1}$ by (\ref{eq:04:05})\;
Convert $\mathbf{Z}^{t+1}$ to the spatial-frequency domain by AD2SF module with zero padding\;
Update penalty factor $\rho^{t+1}=\alpha\rho^{t}$;	}
return $\mathbf{Z}^{N+1}$
\end{algorithm}		
	
\par
Subproblem (\ref{eq:04:03b}) can be solved by the DL-based denoiser in the subsection \ref{deep PnP prior}. Algorithm \ref{alg:04:03} shows the plug-and-play antenna extrapolation, i.e., PPPAE. The results obtained from the spline interpolation are used as the initialization. The DL model in the PPPAE can be used for any antenna selection patterns, while the existing DL-based antenna extrapolation methods \cite {49} can only deal with fixed antenna selections and several DL models need to be trained and stored to work on different antenna selection patterns.
	
\subsection{CSI Feedback via Deep PnP prior}
\par
This subsection describes the proposed algorithm PPPCF for the CSI feedback task. According to (\ref{eq:03:03}), (\ref{eq:04:01}), and (\ref{eq:04:02}), subproblem (\ref{eq:04:03a}) under the CSI feedback task can be expressed as
	\begin{equation}\label{eq:03:06}
		\bar{\mathbf{h}}^{t+1}:=\arg\min_{\bar{\mathbf{h}}}\left\|\mathbf{y}-\mathbf{A}\bar{\mathbf{h}}\right\|_{2}^{2}+\rho\left\|\mathbf{z}^{t}-\bar{\mathbf{h}}\right\|_{2}^{2}.
	\end{equation}
For this unconstrained optimization problem, the closed-form solution is given by
\begin{equation}\label{eq:03:07}		
\bar{\mathbf{h}}^{t+1} = \left(\mathbf{A}^{\text{T}}\mathbf{A}+\rho\mathbf{I} \right)^{-1} \left(\mathbf{A}^{\text{T}}\mathbf{y}+\rho\mathbf{z}^{t} \right),
\end{equation}
where $\mathbf{I}$ is the identity matrix. Unlike (\ref{eq:04:09}) and (\ref{eq:04:13}), the computational complexity of (\ref{eq:03:07}) is large due to the inverse of the high-dimensional matrix. In order to reduce the computational complexity of (\ref{eq:03:07}), we simplify the operation by performing singular value decomposition of the linear mapping matrix $ \mathbf{A} $ in (\ref{eq:03:07}), and the simplified closed-form solution can be expressed as
\begin{equation}\label{eq:03:13}
\resizebox{0.85\linewidth}{!}{$
			\displaystyle
			\bar{\mathbf{h}}^{t+1} = \left(\mathbf{V}^{\text{T}}\begin{bmatrix}
				\frac{1}{1+\rho}\mathbf{I} &\mathbf{0} \\
				\mathbf{0} &\frac{1}{\rho}\mathbf{I}
			\end{bmatrix}\mathbf{V}\right) \left(\mathbf{A}^{\text{T}}\mathbf{y}+\rho\mathbf{z}^{t}\right),
		$}
\end{equation}
where $ \mathbf{V}^{\text{T}} $ is the right unitary matrix obtained after the SVD of linear mapping matrix $ \mathbf{A} $. The derivation of (\ref{eq:03:13}) is given in Appendix \ref{proof4}. Although penalty factor $\rho$ is constantly updated iteratively, we avoid computing the matrix inverse for each iteration.
	
	
\par
Unlike channel estimation and antenna extrapolation, the iterative process of CSI feedback is performed on the angular-delay domain. Therefore, the denoiser used for the subproblem (\ref{eq:04:03b}) of CSI feedback does not require SF2AD and AD2SF modules. Algorithm \ref{alg:04:04} describes the plug-and-play CSI feedback, i.e., PPPCF. Similarly, PPPCF can be used for any compression ratio since the denoiser is decoupled from the linear mapping that determines the compression ratio.

\begin{algorithm}[tb!]
\SetAlgoLined
\LinesNumbered
\DontPrintSemicolon
\caption{PPPCF.}
\label{alg:04:04}
\KwIn{The compressed CSI vector $\mathbf{y}$, linear mapping matrix $\mathbf{A}$, regularization parameters $\lambda$, penalty factor $\rho$, scaling factor $\alpha$, number of iterations $N$}
\KwOut{Reconstructed CSI vector $\hat{\bar{\mathbf{h}}}$}
Initialization, get $\mathbf{z}^{1}$\;
\For{$t = 1:N$}{
Compute $\bar{\mathbf{h}}^{t+1}$ by (\ref{eq:03:13})\;
Compute $\mathbf{z}^{t+1}$ by (\ref{eq:04:05})\;
Update penalty factor $\rho^{t+1}=\alpha\rho^{t}$;	}
return $\mathbf{z}^{N+1}$
\end{algorithm}	

\section{Experimentel Results}
\label{Experimentel Results}
\par
This section evaluates the performance of the proposed method for the three tasks of channel estimation, antenna extrapolation, and CSI feedback.
	
\subsection{Data Generation and Network Training}
\par
\begin{figure}[!tb] 
\centering
\includegraphics[scale=0.05]{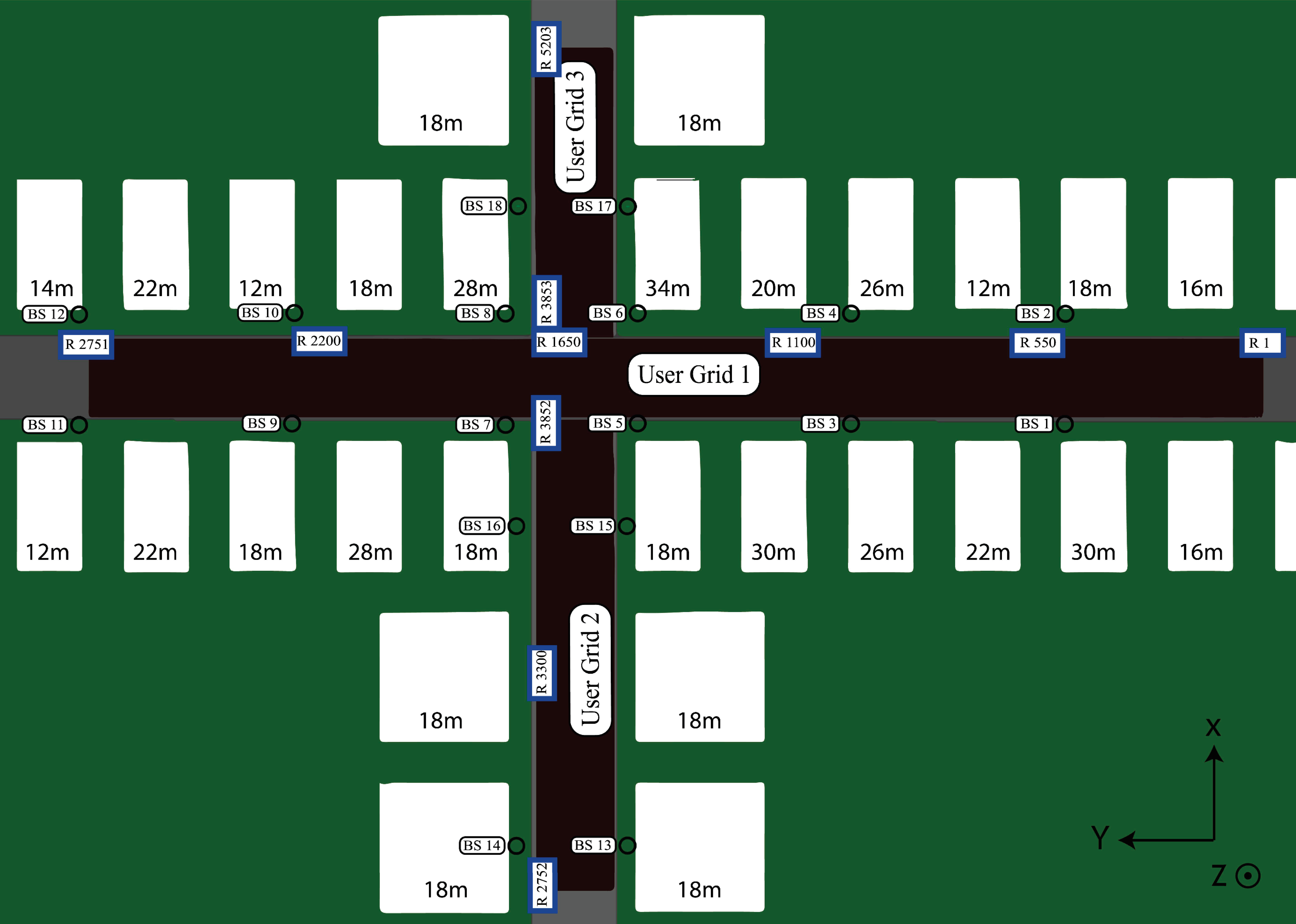}
\caption{Schematic diagram of O1\_28 scenario\cite{100}.}\label{fig:04:03}
\end{figure}

The dataset used in our experiments is generated by the DeepMIMO channel generation platform \cite{100}, which constructs MIMO channel data from accurate ray-tracing data obtained by the 3D ray-tracing software Remcom Wireless InSite\footnote{https://www.remcom.com/wireless-insite-em- propagation-software} and different ray tracing scenarios and parameter sets can be used to achieve accurate definition and reproduction of the dataset. As shown in Fig.~\ref{fig:04:03}, the scene of this experimental data set is the outdoor scene "O1\_28" with center frequency 28 GHz.
	
\par
Referring to the setup in the literature \cite{49}, only BS3 is activated in this experiment, as shown in Fig.~\ref{fig:04:03}. The BS is equipped with a ULA\footnote{The proposed method can also be applied to other types of antennas. Here we consider the ULA model for simplicity and fair comparison with other methods in literature.} of 32 antennas, and each UE has a single antenna. The system bandwidth is 200 MHz and the number of paths is 5. Users are placed in 3 uniform x-y grids, and this experiment activates the second user grid located in the south side of the street in Fig.~\ref{fig:04:03}, which contains 1101 rows, each row has 181 users. The spacing of this uniform x-y grid is 20cm. 150,000 CSI samples were randomly selected as the dataset. The other parameters are shown in Table~\ref{table:04:01}.
	
\begin{table}[!tb]
\centering
\renewcommand\arraystretch{1.5}
\caption{DeepMIMO dataset simulation settings.}
\begin{tabular}{c|c}
\hline
\textbf{Parameters} & \textbf{Values} \\
\hline
Scenarios & O1\_28 \\
Active BS & BS3 \\
Active UEs   & User Grid 2   \\
Number of BS antennas  & 32 \\
Number of UE antennas & 1 \\
UE sampling factor & 0.8 \\
Center frequency   & 28 GHz \\
Number of OFDMs & 1024 \\
OFDM sampling factor & 1 \\
OFDM limit  & 256 \\
Number of paths  & 5 \\
\hline
\end{tabular}%
\label{table:04:01}
\end{table}
	
\par
The generated CSI samples are separated into the training dataset of 100,000 samples and the validation data set of 30,000 samples, which are used to train the denoising network, and the testing dataset of 20,000 samples, which is used to evaluate the proposed method. First, we normalize the power of the spatial-frequency domain CSI, and then add a random Gaussian noise uniformly generated in the range of $0\sim40$dB, which produces the noisy spatial-frequency domain CSI. Next, the clean spatial-frequency domain CSI and the noisy spatial-frequency domain CSI are transformed to the angular-delay domain via DFT, and then truncated \cite{56,60,57,8951228,61,9585309,9373670}, i.e., only the first 32 rows of data are retained to obtain the final dataset. Each training sample contains the clean angular-delay domain CSI, the noisy angular-delay domain CSI and the corresponding noise variance.
	
\par
In the training phase, the batch size is set to 128 and the Adam optimizer is used. The denoising network is trained for a total of 200 epochs to converge, and the learning rate is first set to $10^{-4}$ for initialization. When the loss is not reduced within 20 epochs, the learning rate will decrease by half, and the lower limit of the learning rate is set to $10^{- 7}$. In the evaluation phase, the remaining 20,000 samples are used to evaluate different performance of the proposed method and other comparison methods for different tasks. The proposed method is stopped after 10 iterations. Regularization parameter $\lambda$, penalty factor $\rho$, and scaling factor $\alpha$ can be fine-tuned and set to 0.5, 0.1, and 1.5, respectively\footnote{The source code of this paper has been publicly uploaded to https://github.com/wc253/PNPMT}.
	
\subsection{Performance of Channel Estimation}
This subsection evaluates the PPPCE in terms of both convergence and channel estimation accuracy. The metric for assessing the channel estimation accuracy is the normalized mean-squared error (NMSE), which measures the difference between the recovered downlink frequency domain CSI, $\hat{\mathbf{H}}$, and the original downlink CSI, $\mathbf{H}$, and can be expressed as
\begin{equation}\label{eq:03:18}
\text{NMSE}=\mathbb{E}\left( \frac{\left\|\hat{\mathbf{H}}-\mathbf{H} \right\|_{F}^{2}}{\left\|\mathbf{H} \right\|_{F}^{2}}\right) .
\end{equation}

\subsubsection{Convergence}
Recent theoretical work has analyzed the convergence of the PnP prior algorithms. Sreehari et al. present sufficient conditions that ensure convergence of the PnP prior approach \cite{7542195}. Specifically, it requires the denoising operator to be a proximal mapping, which holds if it is nonexpansive and its subgradient is a symmetric matrix. Subsequent research provides the convergence guarantee for various specific conditions, e.g., bounded denoisers~\cite{7744574}, continuous denoisers \cite{8237460}. Generally speaking, the convergence guarantee of the PnP prior algorithms requires some strong assumptions on the denoiser, which usually do not hold in DL based denoisors, but extensive experimental results show that DL based denoisors work well in the PnP prior framework \cite{80,81,82}. This subsection shows the empirical convergence of PPPCE on the DeepMIMO dataset.
	
	\begin{figure}[!tb] 
		\centering
		\includegraphics[scale=0.6]{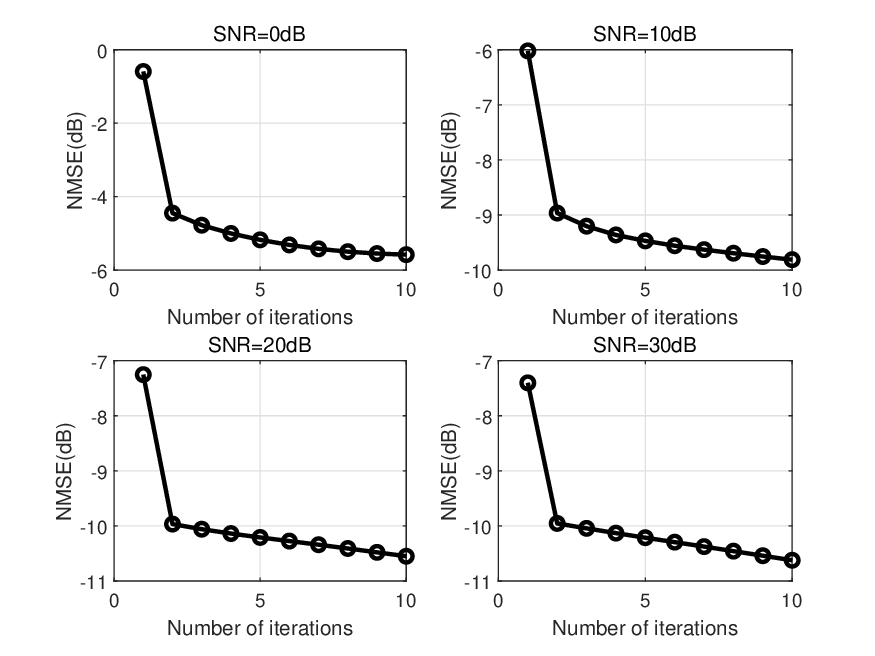}
		\caption{The convergence of PPPCE with 128 pilots.}
		\label{fig:04:04}
	\end{figure}
	
	\begin{figure}[!tb] 
		\centering
		\includegraphics[scale=0.6]{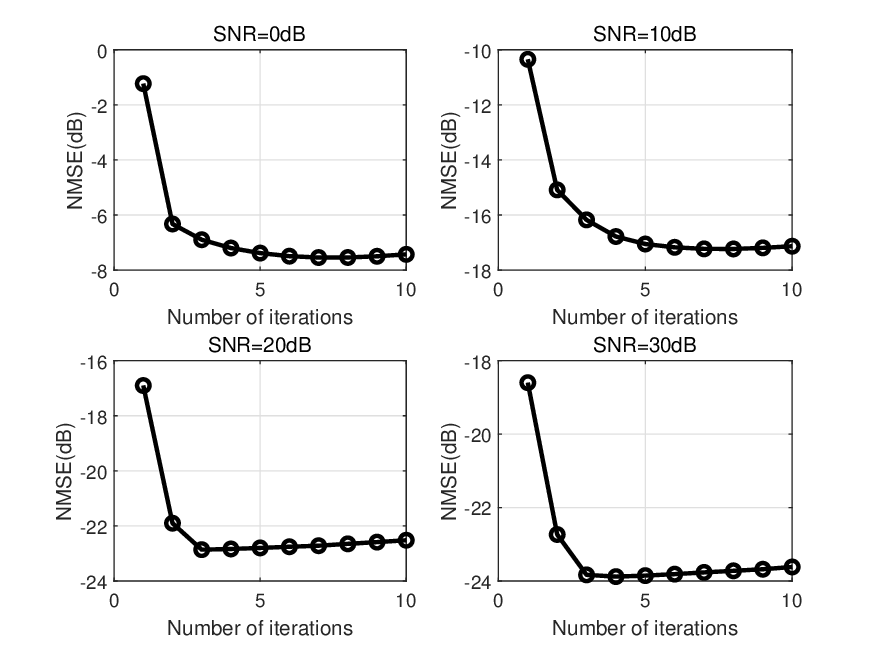}
		\caption{The convergence of PPPCE with 256 pilots.}
		\label{fig:04:05}
	\end{figure}
	
\par
The NMSE convergence curves of PPPCE on different samples with SNR of 0, 10, 20 and 30 dB are shown in Fig.~\ref{fig:04:04} and Fig.~\ref{fig:04:05} for pilot numbers of 128 and 256, respectively. It can be seen that the NMSE of PPPCE with different numbers of pilots all drops quickly in the first several iterations and then the decrease slows down in the subsequent iterations. Therefore, a few iterations would be sufficient to achieve good performance. It should be noted that the proposed deep PnP prior framework implicitly exploits the DL-based denoising operator as the regularizer in (\ref{eq:04:04}), which cannot generally be expressed as a proximal mapping. Therefore, the proposed algorithm does not seek the minimization of an explicit objective function. In Fig.~\ref{fig:04:05}, the NMSE of PPPCE first decreases significantly and then increases slightly as the number of iterations increases.
	
\subsubsection{Channel Estimation Accuracy}
\par
Two classical channel estimation methods, i.e., LS and LMMSE \cite{103}, and two DL-based channel estimation methods, i.e., ChannelNet \cite{34} and ReEsNet \cite{103}, are considered for comparison. It is worth noting that multiple models of ChannelNet and ReEsNet are trained for different pilot patterns. In contrast, the DL model of PPPCE is only for denoising, so a common DL model can be applied for all pilot patterns. To demonstrate that PPPCE is applicable to different pilot patterns, four pilot patterns A, B, C, and D are considered. 128 pilots with the pilot spacing 64 are considered for patterns A and B, and 256 pilots with the spacing 32 are considered for patterns C and D. Different patterns with the same spacing differ in the pilot positions. In addition, ChannelNet and ReEsNet use the dataset of hybrid SNRs, which is also used to train the denoiser of PPPCE.
	
\begin{figure}[!tb] 
		\centering
		\includegraphics[scale=0.6]{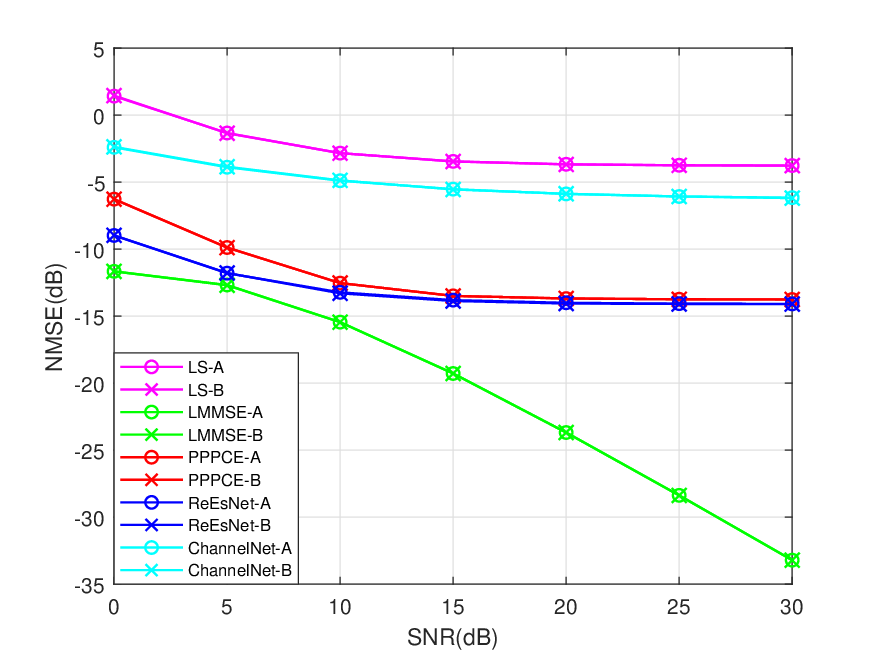}
		\caption{Channel estimation accuracy of different methods with pilot number equal to 128.}
		\label{fig:04:06}
	\end{figure}
	
	\begin{figure}[!tb] 
		\centering
		\includegraphics[scale=0.6]{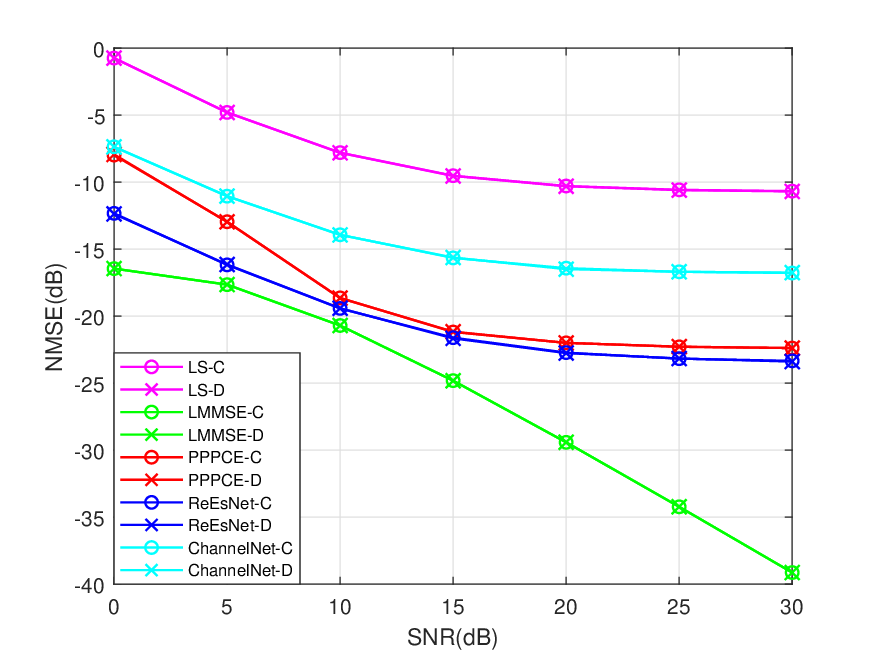}
		\caption{Channel estimation accuracy of different methods with pilot number equal to 256.}
		\label{fig:04:07}
	\end{figure}
	
\par
Fig.~\ref{fig:04:06} and Fig.~\ref{fig:04:07} show the channel estimation accuracy for 128 and 256 pilots, respectively. From the figure, it is evident that the LMMSE method with the highest accuracy is an upper bound, as the estimated results of the LMMSE approach the actual channel under a higher SNR. However, the real channel matrix that the method relies on is not available in reality. The proposed method is primarily compared to other estimation techniques.
The performance of PPPCE is better than that of the ChannelNet and is very close to that of the ReEsNet. However, the ChannelNet and ReEsNet can be only used for channel estimation task in a specific pilot pattern. The DL model in PPPCE is suitable for three different tasks and different scenarios, e.g., the channel estimation task with different pilot patterns, the antenna extrapolation task with different antenna selection patterns, and the CSI feedback task with different compression ratios and quantization bits.

\subsection{Performance of Antenna Extrapolation}
\par
This subsection evaluates the performance of PPPAE.
	
\subsubsection{Convergence}
\par
\begin{figure}[!tb] 
\centering
\includegraphics[scale=0.6]{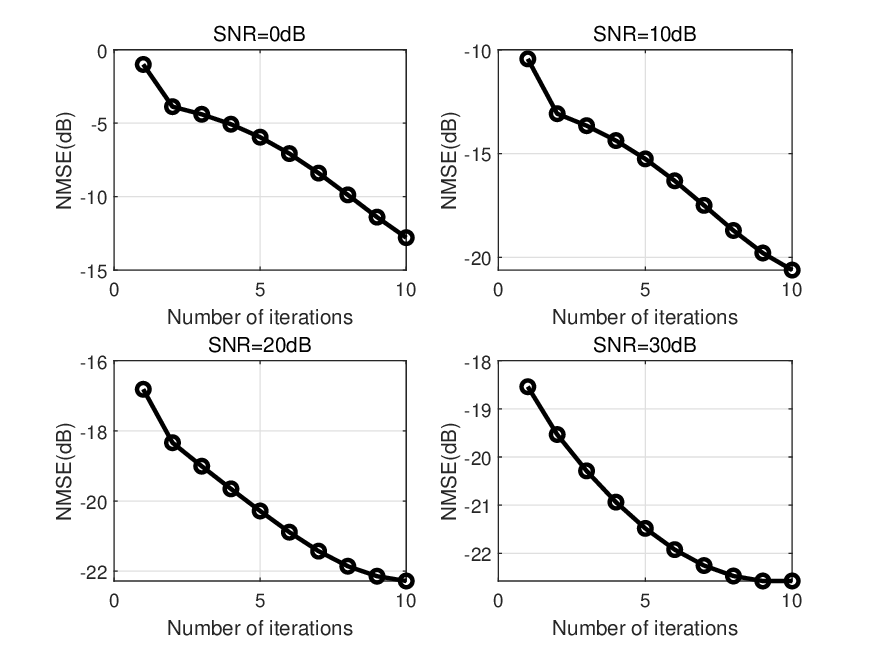}
\caption{The convergence of PPPAE.}
\label{fig:04:08}
\end{figure}

Fig.~\ref{fig:04:08} shows the convergence curve of the NMSE of the PPPAE with 16 selected antennas and SNRs of 0, 10, 20 and 30 dB. From the figure, the NMSE decreases and converges at all the cases with different SNRs. Unlike PPPCE, the NMSE of PPPAE decreases slowly, and more iterations are required to achieve good performance in general.

	
\subsubsection{Antenna Extrapolation Accuracy}
\par
In the antenna extrapolation task, the compared methods include spline interpolation, i.e., RBF, and DL-based ADEN \cite {49} designed specifically for antenna extrapolation. Since ADEN consists of only fully connected layers with a large number of parameters, called FNN-ADEN in our paper, we further design a CNN-based network according to ADEN, namely CNN-ADEN. Note that FNN-ADEN and CNN-ADEN have fixed antenna selection patterns during training, so several DL models need to be trained for different antenna selection patterns. PPPAE is used for denoising, so it is not affected by the different antenna selection patterns.
	
\par
\begin{figure}[!tb] 
\centering
 \includegraphics[scale=0.6]{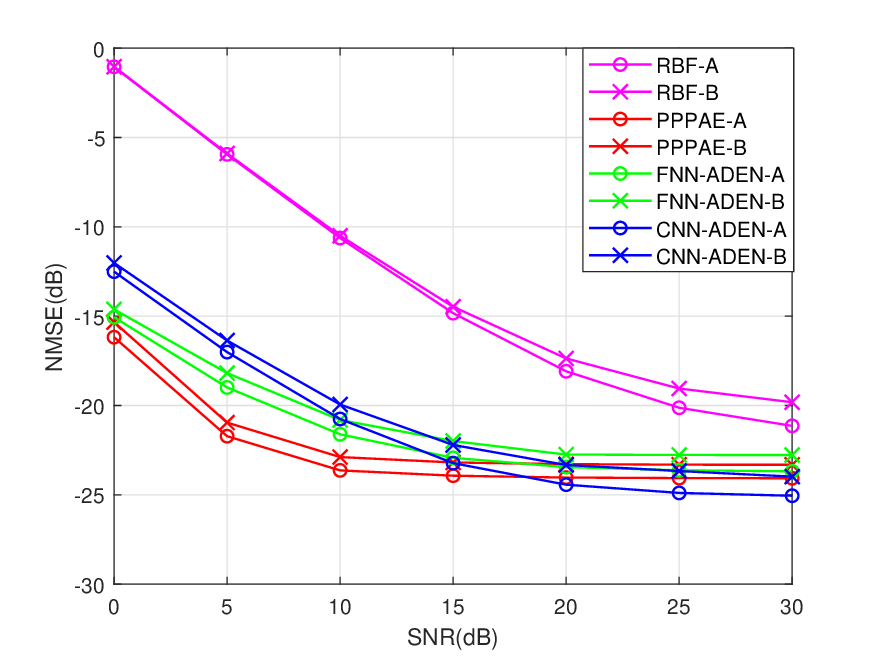}
\caption{Antenna extrapolation accuracy of different methods.}
\label{fig:04:09}
\end{figure}

Similar to the channel estimation task, two antenna selection patterns with 16 antennas are designed, namely A and B, which indicate the selection of odd-numbered and even-numbered antennas. The antenna extrapolation accuracy of each method is shown in Fig.~\ref{fig:04:09}. The accuracy of the antenna extrapolation task differs with distinct antenna selection patterns. Furthermore, the accuracy of PPPAE is better than that of other methods at low SNRs. The DL model used in PPPAE and PPPCE is the same. The proposed approach enables the reuse of DL model for multi-tasks.

\subsection{Performance of CSI Feedback}
\par
This subsection evaluates the performance of PPPCF. In addition to the NMSE, another metric to evaluate the CSI feedback accuracy is the cosine similarity (CoS), which measures the quality of the beamforming vector, given as
\begin{equation}\label{eq:03:19}
\text{CoS}=\mathbb{E}\left( \frac{1}{N_s}\sum_{i=1}^{N_s}\frac{\left|\hat{\mathbf{h}}_{i}^{\text{H}}\mathbf{h}_{i} \right|}{\left\|\hat{\mathbf{h}}_{i} \right\|_{2}\left\|\mathbf{h}_{i} \right\|_{2}}\right),
\end{equation}
where $\hat{\mathbf{h}}_{i}$ and $\mathbf{h}_{i}$ denote the recovered channel vector and the original channel vector of the $i$th subcarrier, respectively.

\begin{figure*}[!t] 
\centering
\includegraphics[scale=0.6]{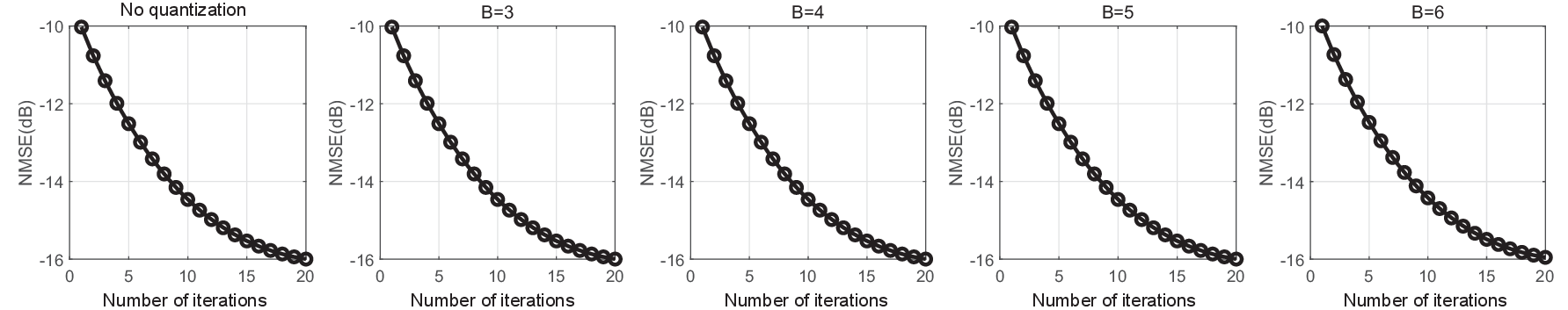}
\caption{The convergence of PPPCF.}
\label{fig:04:12}
\end{figure*}

\subsubsection{Convergence}
According to (\ref{eq:03:01}), the compressed CSI, $\mathbf{A}\bar{\mathbf{h}}$, needs to be quantized for transmission. The quantization error is modeled as the noise term in (\ref{eq:03:01}). Fig.~\ref{fig:04:12} shows the empirical convergence of the PPPCF with CR 1/8. There are five subplots, from left to right, representing different numbers of quantization bits. The results demonstrate the convergence of PPPCF in all cases.
	
\subsubsection{CSI Feedback Accuracy}	
In the CSI feedback task, the proposed method is compared with the CS-based method, namely TVAL3\cite{83}, and CsiNet+\cite{60}, which is based on a two-sided CSI feedback framework that uses the DL model at both the UE side and the BS side.
	
\par
Table \ref{table:04:02} shows the CSI feedback accuracy of different methods under the DeepMIMO dataset. Five different CRs, i.e., 1/4, 1/8, 1/16, 1/32 and 1/64, are considered.  "B" in Table \ref{table:04:02} denotes the number of quantized bits. According to the experimental results, the feedback accuracy of PPPCF is better than TVAL3 in all cases. When CR is 1/4, its feedback accuracy is close to CsiNet+ and even better under some quantized bits. In the remaining cases, the CsiNet+ achieves the best accuracy, which is understandable because the DL-based encoder at the UE side can extract more compact features than simple linear projections, especially when the CR is relatively low or the CSI structure is simpler. Moreover, the sensitivity of PPPCF to quantization bits is much lower than TVAL3 and CsiNet+. At CRs of 1/4, 1/8, and 1/16, the NMSEs increase by 1.44 dB, 1.84 dB, and 2.44 dB for CsiNet+, 4.48 dB, 1.49 dB, and 0.21 dB for TVAL3, and 0.07 dB, 0.24 dB, and 0.14 dB for PPPCF, respectively. This phenomenon is more significant at high CR, which indicates that PPPCF is more robust to the error caused by quantization.
	
\begin{table*}[!tb]
\renewcommand{\arraystretch}{2}
\centering
\caption{Comparison of NMSE (dB) and GoS Performance in the DeepMIMO dataset}
\begin{tabular}{c|c|ccccc}
			\hline
			\textbf{Method} & \diagbox{\textbf{B}}{\textbf{CR}} & 1/4 & 1/8 & 1/16 & 1/32 & 1/64 \\
			\hline
			\multirow{5}{*}{TVAL3} & 3     & -8.86/0.84 & -3.18/0.64 & -1.20/0.49 & -0.41/0.32 & -0.12/0.21 \\
			& 4     & -11.35/0.91 & -4.10/0.71 & -1.37/0.54 & -0.41/0.32 & -0.12/0.21 \\
			& 5     & -12.76/0.94 & -4.41/0.74 & -1.40/0.55 & -0.42/0.33 & -0.12/0.21 \\
			& 6     & -13.34/0.96 & -4.56/0.75 & -1.41/0.56 & -0.42/0.33 & -0.12/0.21 \\
			& no quantization    & -13.62/0.96 & -4.67/0.75 & -1.41/0.56 & -0.42/0.33 & -0.12/0.22\\
			\hline
			\multirow{5}{*}{CsiNet+} & 3     & -17.19/0.99 & -16.25/0.99 & -15.31/0.99 & -14.01/0.98 & -13.21/0.98 \\
			& 4     & -18.05/0.99 & -17.39/0.99 & -16.84/0.99 & -16.11/0.99 & -15.43/0.99 \\
			& 5     & -18.48/0.99 & -17.87/0.99 & -17.52/0.99 & -17.13/0.99 & -16.58/0.99 \\
			& 6     & -18.63/0.99 & -18.09/0.99 & -17.75/0.99 & -17.54/0.99 & -16.95/0.99 \\
			& no quantization   & -18.72/0.99 & -18.20/0.99 & -18.07/0.99 & -17.42/0.99 & -17.18/0.99 \\
			\hline
			\multirow{5}{*}{PPPCF} & 3     & -18.40/0.99 & -13.44/0.98 & -8.79/0.94 & -4.29/0.82 & -1.38/0.55 \\
			& 4     & -18.45/0.99 & -13.59/0.98 & -8.84/0.95 & -4.57/0.84 & -1.50/0.57 \\
			& 5     & -18.46/0.99 & -13.65/0.98 & -8.91/0.95 & -4.70/0.84 & -1.61/0.59 \\
			& 6     & -18.47/0.99 & -13.67/0.98 & -8.92/0.95 & -4.73/0.84 & -1.63/0.59 \\
			& no quantization   & -18.47/0.99 & -13.68/0.98 & -8.93/0.95 & -4.75/0.84 & -1.64/0.59 \\
			\hline
		\end{tabular}%
		\label{table:04:02}%
	\end{table*}%

\subsection{Model Storage}
\par
This subsection discusses the model storage overhead of the other DL-based methods and the proposed method. Table \ref{table:04:03} gives the number of model parameters required to implement each DL-based method for different communication tasks.
For the compared method using different models for different tasks, the total model storage overhead includes the amount of model parameters of a channel estimation model, an antenna extrapolation model and a CSI feedback model. In the proposed method, only one common DL model is needed to be deployed for three different tasks. Therefore, the model storage overhead of the proposed method is much less than that of the compared method. Specifically, for the channel estimation task with different pilot patterns, ChannelNet and ReEsNet have 1255.8K and 340.9K parameters, respectively. For the antenna extrapolation task with different antenna selection patterns, FNN-ADEN and CNN-ADEN have 27300.9K and 365.7K parameters, respectively. For the CSI feedback task with different CRs, CsiNet+ has 4095.4K parameters. In contrast, the DL model in the proposed method requires only 175.2K parameters, which is only 3.6\% of the model parameters in the methods compared to handle all the three tasks.

	
	\begin{table}[!tb]
		\renewcommand{\arraystretch}{2}
		\centering
		\caption{Comparison of model storage overhead}
		\begin{tabular}{c|c|c}
			\hline
			\textbf{Task} & \textbf{Method} & \textbf{Params} \\
			\hline
			\multirow{2}{*}{Channel Estimation} & ChannelNet & 1255.8K \\
			& ReEsNet & 340.9K \\
			\hline
			\multirow{2}{*}{Antenna Extrapolation} & FNN-ADEN & 27300.9K \\
			& CNN-ADEN & 365.7K \\
			\hline
			CSI Feedback & CsiNet+ & 4095.4K \\
			\hline
			All The Three Tasks   & PPPCE, PPPAE, PPPCF & 175.2K \\
			\hline
		\end{tabular}
		\label{table:04:03}
	\end{table}
	
	\par
	
	

\section{Conclusions}
\label{sec:Conclusions}
In this paper, we propose a novel multi-task method for channel reconstruction in massive MIMO systems, where a single DL model can be used for multiple channel reconstruction tasks. Three downlink channel reconstruction tasks are investigated, including channel estimation, antenna extrapolation, and CSI feedback. Using deep PnP prior, we developed PPPCE, PPPAE, and PPPCF for different tasks, which share a common DL model. The DL model does not require labeled dataset for different tasks, and the trained model can be reused in all tasks, which significantly reduces model training costs and model storage overhead. Extensive experiments show the advantages of the proposed method.

	
	
	\appendices
	\section{Proof of the global optimal solution of (\ref{eq:04:06})}
	\label{proof1}
	\par
	(\ref{eq:04:06}) is an unconstrained optimization problem with the optimization terms $ \left\|\mathbf{Y}_{p}-\mathcal{P}(\mathbf{H}) \circ \mathbf{X}_{p}\right\|_{F}^{2} $ and $ \left\|\mathbf{Z}^{t}-\mathbf{H}\right\|_{F}^{2} $, which is a least squares optimization problem of the variable $ \mathbf{H} $. There exists a global optimal solution and it makes (\ref{eq:04:07}) hold
	\begin{equation}\label{eq:04:07}		
		\frac{\partial\left(\left\|\mathbf{Y}_{p}-\mathcal{P}(\mathbf{H}) \circ \mathbf{X}_{p}\right\|_{F}^{2}+\rho\left\|\mathbf{Z}^{t}-\mathbf{H}\right\|_{F}^{2}\right) }{\partial\mathbf{H}}=0.
	\end{equation}
	
	\par
	Since the transmitted signal at the non-pilot position is zero, the numerator on the left side of (\ref{eq:04:07}) can be divided into two sets $\mathbb{P}$ and $\bar{\mathbb{P}}$ according to the pilot and non-pilot positions, which is shown as follows
	\begin{equation}\label{eq:04:08}
		\resizebox{0.85\linewidth}{!}{$
			\displaystyle
			\begin{aligned}
				&\frac{\partial\left(\left\|\mathbf{Y}_{p}-\mathcal{P}(\mathbf{H}) \circ \mathbf{X}_{p}\right\|_{F}^{2}+\rho\left\|\mathbf{Z}^{t}-\mathbf{H}\right\|_{F}^{2}\right) }{\partial\mathbf{H}}\\
				&=\frac{\partial\left(\sum_{\left(i,\ j\right)\in \mathbb{P}}\left(y_{ij}-h_{ij}x_{ij}\right)^{2}+\rho\sum_{\left(i,\ j\right)\in \mathbb{P}}\left(h_{ij}-z_{ij}^{t}\right)^{2}+\rho\sum_{\left(i,\ j\right)\in \bar{\mathbb{P}}}\left(h_{ij}-z_{ij}^{t}\right)^{2} \right) }{\partial\mathbf{H}},
			\end{aligned}$}
	\end{equation}
	where $h_{ij}$, $x_{ij}$, $y_{ij}$, and $z_{ij}^{t}$ denote the downlink spatial-frequency domain CSI, $ \mathbf{H} $, the transmitted pilots $ \mathbf{X}_{p} $, the received pilots $ \mathbf{Y}_{p} $, and the result of the denoiser in the $t$th iteration $ \mathbf{Z}^{t} $ corresponds to the value of $\left(i,\ j\right)$ position. If (\ref{eq:04:08}) is equal to zero, the closed-form solution of (\ref{eq:04:06}) is obtained by (\ref{eq:04:019}).
	\begin{equation}\label{eq:04:019}
		h_{ij}^{t+1}=\begin{cases}
			\frac{x_{ij}y_{ij}+\rho z_{ij}^{t}}{x_{ij}^{2}+\rho},&\left(i,\ j\right)\in \mathbb{P}\\
			z_{ij}^{t},&\left(i,\ j\right)\in \bar{\mathbb{P}}.
		\end{cases}
	\end{equation}

	\section{Proof of the global optimal solution of (\ref{eq:04:10})}
	\label{proof2}
	\par
	(\ref{eq:04:10}) is also an unconstrained optimization problem and its global optimal solution makes (\ref{eq:04:11}) hold
	\begin{equation}\label{eq:04:11}		
		\frac{\partial\left(\left\|\tilde{\mathbf{H}}-\mathcal{A}\left( \mathbf{H} \right)\right\|_{F}^{2}+\rho\left\|\mathbf{Z}^{t}-\mathbf{H}\right\|_{F}^{2}\right) }{\partial\mathbf{H}}=0.
	\end{equation}

	\par
	Since only some of the antennas are selected, the numerator on the left side of (\ref{eq:04:11}) can be divided into two sets $\mathbb{A}$ and $\bar{\mathbb{A}}$ by the set of selected antennas and the set of unselected antennas, which is shown as follows
	\begin{equation}\label{eq:04:12}
		\resizebox{0.85\linewidth}{!}{$
			\displaystyle
			\begin{aligned}
				&\frac{\partial\left(\left\|\tilde{\mathbf{H}}-\mathcal{A}\left( \mathbf{H} \right)\right\|_{F}^{2}+\rho\left\|\mathbf{Z}^{t}-\mathbf{H}\right\|_{F}^{2}\right) }{\partial\mathbf{H}}\\
				&=\frac{\partial\left(\sum_{\left(i,\ j\right)\in \mathbb{A}}\left(\tilde{h}_{ij}-h_{ij}\right)^{2}+\rho\sum_{\left(i,\ j\right)\in \mathbb{A}}\left(h_{ij}-z_{ij}^{t}\right)^{2}+\rho\sum_{\left(i,\ j\right)\in \bar{\mathbb{A}}}\left(h_{ij}-z_{ij}^{t}\right)^{2} \right) }{\partial\mathbf{H}},
			\end{aligned}$}
	\end{equation}
	where $h_{ij}$, $\tilde{h}_{ij}$, and $z_{ij}^{t}$ denote the downlink CSI, $ \mathbf{H} $, the selected set of antennas corresponding to the downlink CSI, $ \tilde{\mathbf{H}}$, and the result of the denoiser in the $t$th iteration $\mathbf{Z}^{t} $ corresponding to the value of $\left(i,\ j\right)$ position, respectively. If (\ref{eq:04:12}) is equal to zero, the closed-form solution of (\ref{eq:04:10}) is obtained by
	\begin{equation}\label{eq:04:23}
		h_{ij}^{t+1}=\begin{cases}
			\frac{\tilde{h}_{ij}+\rho z_{ij}^{t}}{1+\rho},&\left(i,\ j\right)\in \mathbb{A}\\
			z_{ij}^{t},&\left(i,\ j\right)\in \bar{\mathbb{A}}.
		\end{cases}
	\end{equation}


	
	

\section{Proof of (\ref{eq:03:13})}
\label{proof4}
\par
First, we can perform SVD on the linear mapping matrix $\mathbf{A}$.
\begin{equation}\label{eq:03:10}
\mathbf{A}=\mathbf{U} \boldmath{\Sigma} \mathbf{V}^{\text{T}}
\end{equation}
where $\mathbf{U}\in \mathbb{R}^{M \times M}$ and $\mathbf{V}\in \mathbb{R}^{N \times N}$ are unitary matrices. Note that it is explained in section \ref{CSI feedback} that the rows of $\mathbf{A}$ are mutually orthogonal, so the first $M$ columns of $\boldmath{\Sigma} \in \mathbb{R}^{M \times N}$ is a  unit matrix and the rest of the columns are zeros. The partitioned matrix form of $\boldmath{\Sigma}$ is
\begin{equation}\label{eq:03:11}
\boldmath{\Sigma}=\left[\mathbf{I}\ \ \mathbf{0}\right],
\end{equation}
where $\mathbf{I}\in \mathbb{R}^{M \times M}$ is a unit matrix and $\mathbf{0}\in \mathbb{R}^{M \times \left(N-M \right) }$ is an all-zero matrix. Combine (\ref{eq:03:10}) with (\ref{eq:03:11}), then $\mathbf{A}^{\text{T}}\mathbf{A}$ in (\ref{eq:03:07}) is expanded as
\begin{equation}\label{eq:03:12}
\mathbf{A}^{\text{T}}\mathbf{A}=\mathbf{V}\begin{bmatrix}
\mathbf{I} &\mathbf{0} \\
\mathbf{0} &\mathbf{0}
\end{bmatrix}\mathbf{V}^{\text{T}}.
\end{equation}

\par
Substitute (\ref{eq:03:12}) into (\ref{eq:03:07}) and (\ref{eq:03:07}) is further expanded as
	\begin{equation}\label{eq:03:111}
		\begin{aligned}
			\bar{\mathbf{h}}^{t+1} &= \left(\mathbf{A}^{\text{T}}\mathbf{A}+\rho\mathbf{I} \right)^{-1} \left(\mathbf{A}^{\text{T}}\mathbf{y}+\rho\mathbf{z}^{t} \right)\\
			&=\left(\mathbf{V}\begin{bmatrix}
				\mathbf{I} &\mathbf{0} \\
				\mathbf{0} &\mathbf{0}
			\end{bmatrix}\mathbf{V}^{\text{T}}+\rho\mathbf{I} \right)^{-1} \left(\mathbf{A}^{\text{T}}\mathbf{y}+\rho\mathbf{z}^{t}\right)\\
			&=\left(\mathbf{V}\left( \begin{bmatrix}
				\mathbf{I} &\mathbf{0} \\
				\mathbf{0} &\mathbf{0}
			\end{bmatrix}+\rho\mathbf{I}\right) \mathbf{V}^{\text{T}} \right)^{-1} \left(\mathbf{A}^{\text{T}}\mathbf{y}+\rho\mathbf{z}^{t}\right)\\
			&=\left(\mathbf{V}\begin{bmatrix}
				\left(1+\rho \right) \mathbf{I} &\mathbf{0} \\
				\mathbf{0} &\rho\mathbf{I}
			\end{bmatrix}\mathbf{V}^{\text{T}} \right)^{-1} \left(\mathbf{A}^{\text{T}}\mathbf{y}+\rho\mathbf{z}^{t}\right)\\
			&=\left(\mathbf{V}^{\text{T}}\begin{bmatrix}
				\frac{1}{1+\rho}\mathbf{I} &\mathbf{0} \\
				\mathbf{0} &\frac{1}{\rho}\mathbf{I}
			\end{bmatrix}\mathbf{V}\right) \left(\mathbf{A}^{\text{T}}\mathbf{y}+\rho\mathbf{z}^{t}\right).		
		\end{aligned}
	\end{equation}
	
	\balance
	\bibliographystyle{IEEEtran}
	\bibliography{refs}
	
\end{document}